%Paper: gr-qc/9304042
%From: snh@ibm-1.MPA-Garching.MPG.DE (Sean Hayward)
%Date: Thu, 29 Apr 93 13:55:18 +0200

\font\lbf=cmbx10 scaled\magstep2
\font\sm=cmr7

\def\bs{\bigskip}
\def\ms{\medskip}

\def\ni{\noindent}
\def\cl{\centerline}

\def\title#1{\cl{\lbf #1}\ms}

\def\ref#1#2#3#4{#1\ {\it#2\ }{\bf#3\ }#4\par}
\def\refb#1#2#3{#1\ {\it#2\ }#3\par}
\def\ns{\kern-.33333em}
\def\ANY{Ann.\ N.Y.\ Acad.\ Sci.}
\def\APP{Acta Phys.\ Pol.}
\def\CQG{Class.\ Qu.\ Grav.}
\def\PR{Phys.\ Rev.}
\def\PRL{Phys.\ Rev.\ Lett.}

\def\address{\ms\cl{Max-Planck-Institut f\"ur Astrophysik}
		\cl{Karl-Schwarzschild-Stra\ss e 1}
		\cl{8046 Garching bei M\"unchen}
		\cl{Germany}\ms}
\def\me{\ms\cl{\bf Sean A. Hayward}\address}

\def\H{{\cal H}}
\def\L{{\cal L}}
\def\R{{\cal R}}
\def\a{\alpha}
\def\l{\lambda}
\def\s{\sigma}
\def\t{\theta}
\def\o{\omega}
\def\half{{\textstyle{1\over2}}}
\def\I{\int_S\mu}
\def\A{\sqrt{{A\over{16\pi}}}}

\def\ADM{{\sm ADM}\ }
\def\etal{{\it et al\/\ }}

\magnification=\magstep1
\overfullrule=0pt

\title{On the definition of averagely trapped surfaces}
\me
\cl{27th April 1993}
\ms\ni
{\bf Abstract.}
Previously suggested definitions of averagely trapped surfaces
are not well-defined properties of 2-surfaces,
and can include surfaces in flat space-time.
A natural definition of averagely trapped surfaces
is that the product of the null expansions be positive on average.
A surface is averagely trapped in the latter sense if and only if
its area $A$ and Hawking mass $M$ satisfy the isoperimetric inequality
$16\pi M^2 > A$,
with similar inequalities existing for other definitions of quasi-local energy.
\bs\ni
An important concept in General Relativity is the trapped surface
(Penrose 1965), which provides the practical definition of a black hole,
and is the key ingredient in the singularity theorems (Hawking and Ellis 1973).
Hartle and Wilkins (1973) and Bizon and Malec (1989)
have suggested that it may also be useful
to study surfaces which are averagely trapped in some sense,
and various results have been obtained relating the Bizon-Malec definition
to initial data (Malec 1991ab, Zannias 1993, Koc 1993).
The aim of this note is to show that
the Hartle-Wilkins and Bizon-Malec definitions have unacceptable properties,
but that a suitable definition of averagely trapped surfaces does exist.
\ms
Firstly, recall the conventional definitions of trapped surfaces.
Consider a compact orientable spatial 2-surface $S$ with area form $\mu$.
The surface has two future-pointing null normal vectors $n_\pm$,
which are normalised by $n^+_an_-^a=-1$ and defined uniquely
up to the boost transformation $n_\pm\mapsto\l^{\pm1}n_\pm$.
The expansions $\t_\pm$ in these directions are defined by
$\mu\t_\pm=\L_\pm\mu$, where $\L_\pm$ denotes the Lie derivative along $n_\pm$.
Then $S$ is said to be (strictly) {\it trapped} if $\t_+\t_->0$ on $S$.
One may distinguish future and past trapped surfaces by
$\t_\pm<0$ and $\t_\pm>0$ respectively, corresponding to black and white holes.
Similarly, a surface is marginally trapped, or {\it marginal}, if $\t_+\t_-=0$.
\ms
In addition,
one may define a (future) {\it outer trapped} surface by $\t_o\le0$
(Hawking and Ellis 1973), where the expansions $\t_\pm$
have been identified in some way as inner and outer,
$\t_i$ and $\t_o$ respectively.
In general,
there is no way to distinguish the inside and outside of a 2-surface.
The outer trapped surface is well-defined only in particular circumstances
where there is some natural way to assign an interior or exterior
to the 2-surface,
such as in an asymptotically flat space-time,
or for a 2-surface bounding a compact spatial 3-surface,
or for a 2-surface lying on a light-cone.
Some such situation will be assumed where necessary.
\ms
The Bizon-Malec definition of an averagely outer trapped surface is
intended as a weakening of the concept of outer trapped surface,
namely $\I\t_o\le0$,
the idea being that $\t_o$ need not be negative everywhere on $S$,
but only on average.
The problem is that $\t_o$ is not boost-invariant,
since the expansions transform as $\t_\pm\mapsto\l^{\pm1}\t_\pm$ under a boost.
Thus the definition is not a well-defined property of 2-surfaces.
By choosing the boost factor $\l$,
the magnitude of $\t_o$ can be made arbitrarily large or small at any point.
For instance, any surface on which $\t_o$ varies its sign
can be made either averagely trapped or not,
simply by choosing $\l$ to be sufficiently large or small
where $\t_o$ is negative.
In particular, this can be done in flat space-time,
so that such averagely trapped surfaces exist in flat space-time.
\ms
Fixing the boost freedom requires some preferred direction,
which could be provided by a preferred 3-surface passing through $S$.
Hence the Bizon-Malec definition could be rephrased as
a property of spatial 3-surfaces with boundary, rather than of 2-surfaces.
This is at least logically consistent, but still has unacceptable properties:
one 3-surface may be averagely trapped,
and another with the same boundary may not be.
This means that one cannot speak about a region of space-time
being averagely trapped in this sense.
Also, as above, such averagely trapped 3-surfaces exist in flat space-time.
Therefore the concept is not an indicator of gravitational collapse.
\ms
Note that the results which relate these so-called averagely trapped surfaces
to initial data (Bizon and Malec 1989, Malec 1991ab, Zannias 1993, Koc 1993)
all require special assumptions to the effect that
a preferred 3-surface exists, such as a surface of time symmetry.
In such restricted contexts, the problem with the definition is not apparent.
\ms
The Hartle-Wilkins definition of an averagely (future) trapped surface is
intended as a weakening of the concept of (future) trapped surface,
namely $\I\t_+<0$, $\I\t_-<0$.
Again, this is not a well-defined property of 2-surfaces,
since $\t_\pm$ are not boost-invariant.
For instance, for any surface on which one expansion is negative
and the other varies its sign,
the boost freedom can be used
to make the surface either averagely trapped or not.
\ms
Fortunately, there does exist a satisfactory definition
which is an obvious generalisation of that of trapped surfaces.
Namely, a surface is said to be {\it averagely trapped} if $$\I\t_+\t_->0.$$
In words, $\t_+\t_-$ need not be positive everywhere on $S$,
but only on average.
This is a well-defined property of 2-surfaces,
with the combination $\t_+\t_-$ being boost-invariant.
Similarly, an {\it averagely marginal} surface is defined by $\I\t_+\t_-=0$.
\ms
The question remains whether average trapping is a useful concept.
Roughly speaking, one would expect rather strong gravitational fields
to be required to render $\t_+\t_-$ negative over a large part of a surface,
and so one might hope that the presence of such a surface
would indicate a process of irreversible gravitational collapse
resulting in the formation of a genuinely trapped surface.
There seems to be no firm evidence in favour of this,
nor obvious counter-examples.
\ms
One indication that averagely trapped surfaces might not be completely useless
is that they are closely related to measures of quasi-local mass or energy.
Consider the Hawking mass $$M={1\over{8\pi}}\A\I(\R+\t_+\t_-),$$
where $A=\I$ is the area and $\R$ the Ricci scalar of $S$.
By the Gauss-Bonnet theorem, $\I\R=8\pi$ if $S$ has spherical topology,
and hence $S$ is averagely trapped if and only if $$16\pi M^2>A.$$
This is known as the isoperimetric inequality,
which is thought to be relevant to
the cosmic censorship conjecture and the formation of trapped surfaces,
with $M$ replaced by the \ADM mass (Penrose 1973, Gibbons 1984).
In spherical symmetry, the Hawking mass reduces to the standard
spherically symmetric gravitational energy (Misner and Sharp 1964),
and the above result reduces to the statement that
$S$ is trapped if and only if $16\pi M^2>A$.
\ms
A natural definition of quasi-local energy
is obtained from the `2+2' Hamiltonian $\H$ (Hayward 1993a), namely
$$E=-m\int_S\H
={m\over{8\pi}}\I\left(\R+\t_+\t_--\half\s^+_{ab}\s_-^{ab}-2\o_a\o^a\right),$$
where $\s_\pm$ are the shears and $\o$ the normal fundamental form of $S$,
and for convenience the irreducible mass $m$ has been introduced,
$16\pi m^2=A$.
The term involving the shears ensures that $E$ vanishes
for any surface in flat space-time, a property not shared by the Hawking mass.
In this sense, $E$ can be regarded as a modification of the Hawking mass
which provides a more realistic measure of gravitational energy.
Introducing the quasi-local angular momentum $j$ (Hayward 1993b)
and the `alignment' $\a$, defined by
$$j^2={m^4\over{4\pi}}\I\o_a\o^a,
\qquad\a={m\over{16\pi}}\I\s^+_{ab}\s_-^{ab},$$
it follows that $S$ is averagely trapped if and only if
$$E>m-{j^2\over{m^3}}-\a.$$
This inequality can be regarded as a generalisation
of the isoperimetric inequality beyond the spherically symmetric case.
\ms
Penrose (1973) argued that the cosmic censorship hypothesis requires
the isoperimetric inequality to hold for the \ADM mass,
at least in the time-symmetric case.
In this case, $\t_+\t_-\le0$, so that strictly trapped surfaces cannot occur,
though marginal surfaces can.
Similarly, averagely trapped surfaces cannot occur,
and averagely marginal surfaces must be genuinely marginal.
Also, $\o=0$ and $\s^+_{ab}\s_-^{ab}\le0$,
so that marginal surfaces occur if and only if $E=m-\a$,
with $\a\le0$ and $j=0$.
In particular, the existence of a marginal surface requires
$E\ge m$, or $16\pi E^2\ge A$.
Proving the isoperimetric inequality for the \ADM mass
is thus reduced to proving a suitable monotonicity result for $E$ (or $M$).
\ms
The connection with quasi-local energy and the isoperimetric inequality
gives some indication that
averagely trapped surfaces are a natural and perhaps even useful concept.
It seems likely that the occurrence of such surfaces can be related to
initial data using techniques previously applied to ordinary trapped surfaces
(Bizon \etal 1988, 1989, 1990, Brauer and Malec 1992, Zannias 1992,
Flanagan 1992, Malec and \'O Murchadha 1993).
What is really lacking
is a firm connection between averagely trapped surfaces and black holes.
\bs\ni
I would like to thank Uwe Brauer, Edward Malec and Niall \'O Murchadha
for discussions, and the Max-Planck-Gesellschaft for financial support.
\bs
\begingroup
\parindent=0pt\everypar={\global\hangindent=20pt\hangafter=1}\par
{\bf References}\ms
\ref{Bizon P \& Malec E 1989}\PR{D40}{2559}
\ref{Bizon P, Malec E \& \'O Murchadha N 1988}\PRL{61}{1147}
\ref{Bizon P, Malec E \& \'O Murchadha N 1989}\CQG{6}{961}
\ref{Bizon P, Malec E \& \'O Murchadha N 1990}\CQG{7}{1953}
\ref{Brauer U \& Malec E 1992}\PR{D45}{R1836}
\ref{Flanagan E 1992}\PR{D46}{1429}
\refb{Gibbons G W 1984 in}{Global Riemannian Geometry}
{ed Willmore \& Hitchin (Ellis Horwood Ltd)}
\ref{Hartle J B \& Wilkins D C 1973}\PRL{31}{60}
\refb{Hawking S W \& Ellis G F R 1973}{The Large Scale Structure of Space-Time}
{(Cambridge University Press)}
\refb{Hayward S A 1993a}{Quasi-local gravitational energy}{(gr-qc/9303030)}
\refb{Hayward S A 1993b}{General laws of black-hole dynamics}{(gr-qc/9303006)}
\refb{Koc P 1993}
{Condensation of matter and trapped surfaces in quasi-polar gauge}
{(gr-qc/9304017)}
\ref{Malec E 1991a}\APP{B22}{347}
\ref{Malec E 1991b}\PRL{67}{949}
\ref{Malec E \& \'O Murchadha N 1993}\PR{D47}{1454}
\ref{Misner C W \& Sharp D H 1964}\PR{136B}{571}
\ref{Penrose R 1965}\PRL{14}{57}
\ref{Penrose R 1973}\ANY{224}{125}
\ref{Zannias T 1992}\PR{D45}{2998}
\ref{Zannias T 1993}\PR{D47}{1448}
\endgroup
\bye